\documentclass{elsart}

\usepackage{amsmath}
\usepackage{graphicx}
\usepackage{amssymb}

\newcommand{\corrected}{}

\begin{document}

\begin{frontmatter}
\title{Entropic particle transport in periodic channels}

\author[A]{P. S. Burada},
\author[A]{G. Schmid},
\author[A]{P. Talkner},
 \author[A]{P. H\"anggi}
\author[B]{D. Reguera},
\author[B]{J. M. Rub\'i},
\address[A]{Institut f\"ur Physik,
  Universit\"at Augsburg,
  Universit\"atsstr. 1,
  D-86135 Augsburg, Germany}
\address[B]{Departament de F\'isica Fonamental,
  Facultat de F\'isica,
 Universidad de Barcelona,
  Diagonal 647, E-08028 Barcelona, Spain}

\begin{abstract}
The dynamics of Brownian motion has widespread applications
extending from transport in designed micro-channels  up to its
prominent role for inducing transport in molecular motors and
Brownian motors. Here, Brownian transport is studied in micro-sized,
two dimensional  periodic  channels,  exhibiting periodically
varying cross sections. The particles  in addition are subjected to
an external force acting alongside the direction of the longitudinal
channel axis. For a fixed channel geometry, the dynamics of the two
dimensional problem is characterized by a single dimensionless
parameter which is proportional to the ratio of the applied force
and the temperature of the particle environment. In such structures
entropic effects may play a dominant role. Under certain conditions
the two dimensional dynamics can be approximated by an effective one
dimensional motion of the particle in the longitudinal direction.
The Langevin equation describing this reduced, one dimensional
process is of the type of the Fick-Jacobs equation. It contains an
entropic potential determined by the varying extension of the
eliminated  channel direction, and a correction to the diffusion
constant that introduces a space dependent diffusion. Different
forms of this correction term have been suggested before, which we
here compare for a particular class of models. We analyze the regime
of validity of the Fick-Jacobs equation, both by means of analytical
estimates and the comparisons with numerical results for the full
two dimensional stochastic dynamics. For the nonlinear mobility we
find a temperature dependence which is opposite to that known for
particle transport in periodic potentials. The influence of entropic
effects is discussed for both, the nonlinear mobility and the
effective diffusion constant.
\end{abstract}

\end{frontmatter}

\section {Introduction }

The phenomenon of entropic transport is ubiquitous in biological
cells, ion channels,  nano-porous materials, zeolites and
microfluidic devices etched with grooves and chambers. Instead of
diffusing freely in the host liquid phase the Brownian particles
frequently undergo a constrained motion \cite{liu, siwy,
berzhkovski, hille,
zeolites,Chou,kettner,muller,ai2006,Austin,Nixon,Chang}. The
geometric restrictions to the system's dynamics results in entropic
barriers and regulate the transport of particles yielding important
effects exhibiting peculiar properties. The results have prominent
implications in processes such as catalysis, osmosis and particle
separation
\cite{liu,siwy,berzhkovski,hille,zeolites,Chou,kettner,muller,ai2006,Austin,Nixon,Chang}
and, as well, for the noise-induced transport in periodic potential
landscapes that lack reflection symmetry (Brownian ratchet systems)
\cite{BM,PT,RH} or Brownian motor transport occurring  in arrays of
periodically arranged asymmetric obstacles, termed "entropic"
ratchet devices \cite{entropicR}. Motion in these systems can be
induced by imposing different concentrations at the ends of the
channel, or by the presence of external driving forces supplying the
particles with the energy necessary to proceed. The study of the
kinetics of the entropic transport, the properties of transport
coefficients in far from equilibrium situations and the possibility
for transport control mechanisms are pertinent objectives  in the
dynamical characterization of those systems.

Because the role of inertia for the motion of the particles through
these structures can typically be neglected the Brownian dynamics
can safely be analyzed by solving the Smoluchowski equation in the
domain defined by the available free space upon imposing the
appropriate boundary conditions. Whereas this method has been very
successful when the boundaries of the system possess a rectangular
shape, the challenge to solve the boundary value problem in the case
of nontrivial, corrugated domains  represents  a difficult task. A
way to circumvent this difficulty consists in coarsening the
description by reducing the dimensionality of the system, keeping
only the main direction of transport, but taking into account the
physically available space by means of an entropic potential. The
resulting kinetic equation for the probability distribution, the so
called Fick-Jacobs (FJ) equation, is similar in form to the
Smoluchowski equation, but now contains an entropic term. The
entropic nature of this term leads to a genuine dynamics which is
distinctly  different from that observed when the potential is of
energetic origin \cite{Reguera_PRL}. It has been shown that the FJ
equation can provide a very accurate description of entropic
transport in {\corrected channels of varying cross-section}
\cite{Reguera_PRL,Burada_PRE,Ilona}. However, the derivation of the FJ equation
entails a tacit approximation: The particle distribution in the
transverse direction is assumed to equilibrate much faster than in
the main (unconstrained) direction of transport. This equilibration
 justifies the coarsening of the description leading in turn to a simplification
of the dynamics, but raises the question about its validity when an
{\it external force is applied}. To establish the validity criterion
of a FJ description for such biased diffusion in confined media is,
due to the ubiquity of this situation, a subject of primary
importance.

Our objective with this work is to investigate in greater detail the
FJ-approxi\-mation for biased diffusion and to set up a corresponding
criterion describing its regime of validity. We will analyze the
biased movement of Brownian particles in 2D periodic channels of
varying cross section and formulate different criteria for the
validity of such a FJ-description. On the basis of our numerical and
analytical results we recapitulate the striking and sometimes
counterintuitive features \cite{Reguera_PRL}, which arises from
entropic transport and which are different from those observed in
the more familiar case with energetic, metastable landscapes
\cite{hanggi}.

\section{Diffusion in confined systems}
\label{sec:system}

\begin{figure}[t]
  \centering
  \includegraphics{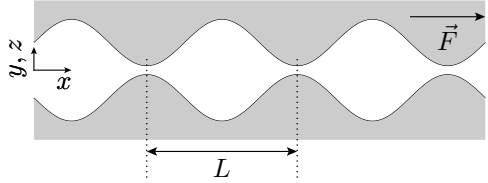}
  \caption{Schematic diagram of a channel confining the
   motion of forced Brownian particles.
   The half-width $\omega$ is a periodic function of
   $x$ with periodicity $L$.}
\label{fig:tube}
\end{figure}

Transport through pores or channels (like the one depicted in
Fig.~\ref{fig:tube})
may be caused by different particle concentrations
maintained at the ends of the channel, or by the application of
external forces acting on the particles. Here we will exclusively
consider the case of force driven transport. The external driving force is
denoted by $\vec{F} = F \vec{e}_x$. It points into the direction of
the channel axis. In
general, the dynamics of a
suspended Brownian particle is overdamped \cite{Purcell} and
well described by the Langevin equation,

\begin{align}
  \label{langevin}
  \eta\, \frac{\mathrm{d}\vec{x}}{\mathrm{d}\tilde{t} } = \vec{F} +
  \sqrt{\eta \, k_\mathrm{B}\,T}\, \vec{\xi}(\tilde{t})\, .
\end{align}

where $\tilde{t}$ is time, $\vec x$ the
position vector of the particle,
 $\eta$ its friction coefficient, $k_{\mathrm{B}}$ the Boltzmann
constant and $T$ the temperature.
The thermal fluctuating forces which the surrounding fluid exerts on
the particle
are modeled by  zero-mean Gaussian white noise
$\vec{\xi}(\tilde{t})$, obeying the fluctuation-dissipation relation
$\langle \xi_{i}(\tilde{t})\,\xi_{j}(\tilde{t}') \rangle = 2\, \delta_{ij}\,
\delta(\tilde{t} - \tilde{t}')$ for $i,j = x,y,z$.

In addition to Eq.~\eqref{langevin}, the full problem is set up by
imposing reflecting boundary conditions at the channel walls. The form
of the channel will be specified below.

To further simplify the treatment of this model we introduce
dimensionless variables. We measure all lengths in units of the period
$L$, i.e.
\begin{equation}
\vec{x} = L \vec{r}\, ,
\end{equation}
where $\vec{r}$ is the dimensionless position vector of the particle.
As unit of time $\tau$ we choose twice the time it takes for the particle to
diffusively cover the distance $L$ which is given by $\tau = L^{2}
\eta /(k_{B}T)$, hence,
\begin{equation}
\tilde{t} = \tau t \, .
\end{equation}
In these dimensionless variables the Langevin equation reads
\begin{align}
  \label{eq:langevin}
  \frac{\mathrm{d}\vec{r}}{\mathrm{d}t } = \vec{f} +
  \vec{\xi}(t)\, .
\end{align}
where $\langle \vec{\xi}(t) \rangle = 0 $ and $\langle \vec{\xi}(t)
\vec{\xi}(t') \rangle = 2 \delta_{i,j} \delta( t- t')$ for $i,j =
x,y,z$ and where the dimensionless force
\begin{equation}
\vec{f} = f \vec{e}_{x} \text{ and} \; f= \frac{LF}{k_{B}T}
\label{f}
\end{equation}
contains the dimensionless parameter $f$ that characterizes the force as
the ratio of the work which it performs on the particle along a distance
of the length of the period  and the thermal energy. 
The corresponding Fokker-Planck equation for the time evolution of
the probability distribution $P(\vec r, t)$ takes the form
\cite{Risken,hanggithomas}:
\begin{subequations}
\label{eq:fp}
\begin{align}
\frac{\partial P(\vec r, t)}{\partial t} = 
- \vec{\nabla} \cdot \vec{J}(\vec{r},t)\, ,
\end{align}
where $\vec J(\vec r, t)$ is the probability current:
\begin{align}
  \label{eq:particlecurrent}
  \vec J(\vec r, t) =
  \left( \vec{f} -
   \vec{\nabla}\right) P(\vec r, t) \, ,
\end{align}
\end{subequations}
\noindent
Note that for channels with similar geometry, which are related
by a scale transformation $\vec{x} \to \lambda \vec{x}$, $\lambda >0$,
the transport properties are determined by
the single dimensionless parameter $f$ which subsumes the period
lengths, the force and the temperature of the surrounding fluid.

The reflection of particles at the channel walls leads to a
vanishing probability current at the boundaries.
Therefore, the boundary conditions at the channel walls are:
\begin{align}
  \label{eq:bc-general}
  \vec J(\vec r, t) \cdot \vec{n}=0\, \quad \vec{r} \in \text{channel wall}.
\end{align}
where $\vec{n}$ denotes the normal vector field at the channel walls.

The boundary of a 2D periodic channel which is mirror symmetric about 
its axis is given by the periodic functions $y= \pm \omega(x)$,
i.e. $\omega(x+1)=\omega(x)$ for all $x$, where $x$ and $y$ are the
 Cartesian components of $\vec{r}$. In this case, the boundary
condition becomes
\begin{equation}
  \label{eq:bc}
  \frac{\mathrm{d} \omega (x)}{\mathrm{d} x}
 \left[ f P(x,y,t) - \frac{\partial P(x,y,t)}{\partial
     x} \right] 
  + \frac{\partial P(x,y,t)}{\partial y} = 0\, ,
  \end{equation}
at $y=\pm \omega (x)$. Except for a straight channel with $\omega =
const $, there are no periodic channel shapes for which an
exact analytical solution of the Fokker-Planck equation \eqref{eq:fp}
with
boundary conditions (\ref{eq:bc}) is known.
Approximate solutions though can be obtained on the basis of an one
dimensional diffusion problem  in an effective potential.
Narrow channel openings, which act as geometric hindrances in the full model,
show up as entropic barriers in this one dimensional approximation
\cite{Reguera_PRL,Burada_PRE,Jacobs,Zwanzig,Reguera_PRE,Percus}.
This
approach  is valid under conditions that will be
discussed  below in some detail.

\section{The Fick-Jacobs approximation}
\label{sec:FJ}

In the absence of an external force, i.e. for $\vec{f}= 0$, it was shown
\cite{Jacobs, Zwanzig, Reguera_PRE,Percus} that
the dynamics of particles in confined structures
(such as that of Fig.~\ref{fig:tube})
can be described approximatively by the FJ equation, with a spatially 
dependent diffusion coefficient:
\begin{align}
\label{eq:fickjacobs}
\frac{\partial P(x,t)}{\partial t}=\frac{\partial}{\partial
  x}\left(D(x)\,\omega(x)\frac{\partial}{\partial
    x} \frac{P}{\omega(x)}\right) \, .
\end{align}
This 1D equation is obtained from the full 2D Smoluchowski 
equation upon the elimination
of the transversal $y$  coordinate assuming fast equilibration in the
transversal
channel direction. Here $P(x,t) = \int_{-\omega(x)}^{\omega(x)}
\mathrm{d}y\,
P(x,y,t)$ denotes the marginal
probability
density along the axis of the channel. We 
note that for three
dimensional channels an analogue approximate Fokker-Planck equation
holds in which the function $\omega(x)$ is to be replaced by
$\pi \omega^{2}(x)$ (area of cross-section).
In the original work by Jacobs \cite{Jacobs} the 1D diffusion coefficient
$D(x)$ is
constant and equals the bare diffusion constant which is unity in the
present dimensionless variables.
Later, Zwanzig \cite{Zwanzig} and Reguera and Rub\'i \cite{Reguera_PRE}
proposed different spatially dependent forms of the 1D diffusion coefficient.

\subsection{Spatially dependent 1D diffusion coefficients}

The 1D diffusion coefficient suggested by Zwanzig results from a systematic
expansion in terms of the gradient of the boundary function
$\omega(x)$. In leading order he obtained $D(x) = 1-\gamma
\omega'(x)^{2} + \ldots \,$, 
{\corrected where $\gamma=1/3$ for the considered 2D structure (for
a 3D structure, $\gamma = 1/2$)} and
the prime denotes the derivative with respect to $x$.
Interpreting this result as the first two terms of a geometric series
Zwanzig proposed his resumed expression for the 1D diffusion coefficient
reading
\begin{align}
  \label{eq:diffusionconst-zwanzig}
  D_{\mathrm{Z}}(x)=\frac{1}{1+\gamma \omega'(x)^{2}}\, .
\end{align}
Reguera and Rub\'i
\cite{Reguera_PRE} put forward a different form of the 1D
diffusion coefficient:
\begin{align}
  \label{eq:diffusionconst}
  D_{\mathrm{RR}}(x)=\frac{1}{(1+\omega'(x)^{2})^{\gamma}} 
\end{align}
which also can be considered as a re-summation of Zwanzig's
perturbational result.
Yet another form of the 1D diffusion coefficient was proposed by Kalinay
and Percus \cite{Percus}. For the biased diffusion discussed here the
results obtained with the Kalinay-Percus diffusion coefficient differ only little
from those obtained by the Reguera-Rub\'i diffusion coefficient. 
Therefore we do not further consider the Kalinay-Percus 
diffusion coefficient.

\subsection{Constant bias along the channel direction}

In the presence of a constant force $F$ along the direction of the
channel the FJ equation~\eqref{eq:fickjacobs} can be recast into the form
\cite{Reguera_PRL,Burada_PRE,Reguera_PRE}:

\begin{align}
\label{eq:fickjacobs_ours}
\frac{\partial P}{\partial t}=\frac{\partial}{\partial
  x}D(x)\left(\frac{\partial P}{\partial
    x}+ \frac{\mathrm{d}A(x)}
    {\mathrm{d}x}P\right)\,
\end{align}
with the dimensionless free energy
$A(x) := E - S = -f\, x -  \ln  \omega(x)$. In physical {\corrected dimensions}
the energy is $\tilde{E} \equiv k_{B}T E = -F \tilde{x}$ ($\tilde{x}
= x L$) and
the dimensional entropic  contribution is $\tilde{S} \equiv k_{B} T S =
k_\mathrm{B} T \,\ln \omega$.
For a periodic channel this free energy assumes the form of a tilted
periodic potential. In the absence of a force the free energy is
purely entropic and
Eq.~(\ref{eq:fickjacobs_ours})
reduces to the FJ equation (\ref{eq:fickjacobs}). On the
other hand, for a straight channel the entropic contribution vanishes
and the particle is solely driven by the external
force.

\subsection {Nonlinear mobility and effective diffusion}

Key quantities of particle transport through periodic channels
are the average {\it particle current}, or equivalently the nonlinear
{\it mobility}, and the {\it effective diffusion coefficient}.
For a particle moving in a one dimensional tilted periodic potential
the heights $\Delta V$ of the barriers separating the potential wells provide an
additional energy scale apart from the work of the force $F L$ and the
thermal energy $k_{B}T$.
Hence, at least two dimensionless parameters, say $\Delta E/(k_{B}T)$
and $FL/(k_{B}T)$ govern
the transport properties of these systems. In contrast, as already
noted in the context of the full 2D model the transport through
channels is governed by the single dimensionless parameter
$f=FL/(k_{B}T)$. This, of course, remains to hold true in the one
dimensional approximation which models the transversal spatial
variation in terms of an entropic potential.

For any non negative force the average particle current in periodic
structures can be obtained from \cite{Reimann_PRL}:
\begin{align}
  \label{eq:parcur}
  \langle \dot{x} \rangle = \langle t(x_{0}\to x_{0}+1)
    \rangle^{-1} \,
\end{align}
where $ t(a\to b)$ denotes the  first time
of a particle which starts at $x=a$ to arrive at $x=b$. The angular
brackets refer to an average over the fluctuating force. Note that the resulting
mean first passage time $\langle t(x_{0}\to x_{0}+1)
    \rangle$ diverges for a vanishing  force
and consequently leads to a vanishing current. A positive force
prevents the particle to make far excursions to the left, hence
leading to a finite mean first passage time as well as a finite current.
Within the one dimensional approximation,
cf. Eq.~\eqref{eq:fickjacobs_ours}, the moments of the first passage
time $ t(a\to b)$ can be determined recursively by means of
\begin{multline}
  \label{eq:fpt}
  \langle t^{n} (a \to b) \rangle =  n \int_{a}^{b} \mathrm{d}x\,
  \frac{1}{D(x)}\, \exp\left(A(x)\right)  \int_{-\infty}^{x}
  \mathrm{d}y\, \exp\left(- A(y) \right)\,\\ \langle t^{n-1}( y \to b )
  \rangle \, .
\end{multline}
For $n=0$ the starting value of the iteration is given by $\langle
t^{0} (a \to b) \rangle = 1$.

The nonlinear mobility $\mu(f)$ is defined by
\begin{equation}
\mu(f) = \frac{\langle \dot{x} \rangle}{f} \, .
\label{mu}
\end{equation}
Using eqs. (\ref{eq:parcur}) and (\ref{eq:fpt}) one can obtain
the following Stratonovich formula for the nonlinear mobility \cite{Reguera_PRL}
\begin{subequations}
  \label{eq:nonlinearmobility}
\begin{align}
  \mu(f) =
\frac{1-\exp(-f)}{ f\: {\displaystyle \int_{0}^{1}}\,
      \mathrm{d}z \, I(z, f)}\, ,
\end{align}
where
\begin{align}
  \label{eq:integrals}
  I(z, f) := \frac{h^{-1}( z)}{D(z)}\,
\exp(- f\, z)\,  \int_{z -
    1}^{z} \mathrm{d}\tilde{z} \, h(\tilde{z})\, \exp(f\, \tilde{z}) \, ,
\end{align}
\end{subequations}
depends on the dimensionless position $z$, the
force $f$ and the shape of the tube given
in terms of the half width $\omega(x)$
and its first derivative.

The effective diffusion coefficient is defined as the asymptotic
behavior of the variance of the position
\begin{equation}
   \label{eq:diffusion-num}
D_{\text{eff}} = \lim_{t \to \infty} \frac{\langle x^{2}(t) \rangle -\langle
  x(t) \rangle^{2}}{2t} \, .
\end{equation}
It is related to the first two moments of the first passage time by
the expression  \cite{Reimann_PRL,Lindner}:
\begin{align}
  \label{eq:diffcoef}
  D_{\mathrm{eff}} =  \frac{\langle
    t^{2}(x_{0}\to x_{0}+1) \rangle - \langle t(x_{0}\to x_{0}+1)
    \rangle^{2}}{2 \:\langle t(x_{0}\to x_{0}+1) \rangle^{3}}\, .
\end{align}
After some algebra it can be transformed to read
\begin{subequations}
  \label{eq:effectivediffusion}
  \begin{align}
    D_\mathrm{eff}= \frac{{\displaystyle \int_{0}^{1}}\mathrm{d}z\,
     {\displaystyle \int_{z-1}^{z}}\mathrm{d}\tilde{z}\,
      {\cal N}(z, \tilde{z}, f)}{\left[{\displaystyle \int_{0}^{1}}\mathrm{d}z\,
        I(z,f)\right]^{3}}\, ,
  \end{align}
  {\corrected where}
  \begin{align}
\label{N}
{\cal N}( z, \tilde{z}, f):=
    \frac{D(\tilde{z})}{h(z)}\,
    \frac{h(\tilde{z})}{D(z)}
    \left[I(\tilde{z},f)\right]^{2} \, \exp(-f\, z + f\,  \tilde{z})\, .
  \end{align}
\end{subequations}

 \subsection{Exact numerics for the 2D channel}
\label{sec:numerics}

The predicted dependence of the average particle current
and the effective diffusion coefficient,
predicted above, was compared  with Brownian
dynamic simulations performed by a numerical integration
of the Langevin equation Eq.~\eqref{eq:langevin}, within the
stochastic Euler-algorithm. The shape of the exemplarily taken 2D
channel is described by
\begin{align}
\omega(x) := a\sin(2\pi x)+ b\,,\label{eq:boundary}
\end{align}
where $b>a$. The sum and difference of the two parameters $a+b$ and
{\corrected $b-a$ give half of the
maximal and the minimal width of the channel}, respectively. Moreover,
$a$ controls the slope of the channel walls which determines the one
dimensional diffusion coefficient $D(x)$.

For the considered channel configuration,
cf. Eq.~\eqref{eq:boundary}, the boundary condition becomes
$\omega(x) = a  \left(\sin(2\pi x) + \kappa \right)$,
where $\kappa = b/a = 1.02$ throughout this paper. For $a$ we chose
values between 1 and $1/2 \pi$. This choice of parameters corresponds to
rather short channels for $a=1$ and a more elongated one for $a=1/2
\pi$. In all cases the width of the widest opening of the channel is
larger by a factor of $100$ than the width at narrowest opening. One
may therefore expect strong entropic effects for these channels.
The particle current and effective diffusion
coefficient were derived
from an ensemble-average  of about $3\cdot 10^{4}$ trajectories:
\begin{align}
  \label{eq:current-num}
  \langle \dot{x}\rangle &=\lim_{t\to \infty}  \frac{\langle x(t)
    \rangle}{t}  \, , 
\end{align}
and Eq.~\eqref{eq:diffusion-num}, respectively.

\label{comparison}
\begin{figure}[t]
\centering
  \includegraphics{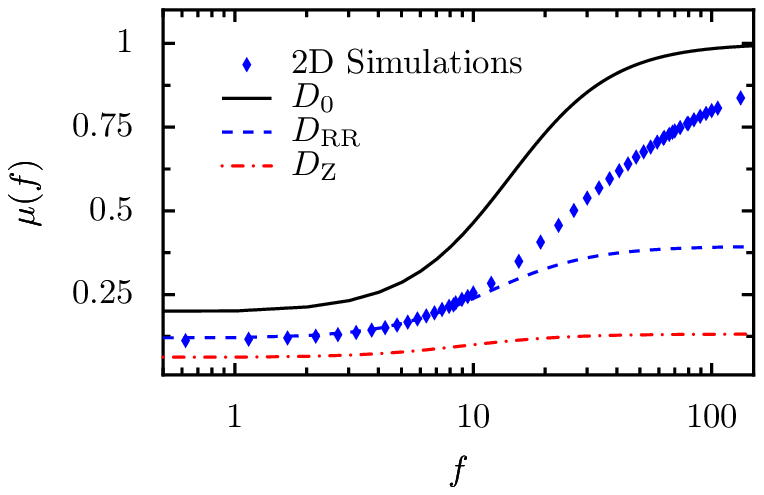}
  \caption{(color online) The dependence of the nonlinear mobility of
    a Brownian particle within a 2D channel with half width
    $\omega(x)=\sin(2 \pi x) + 1.02$ is depicted versus the
    dimensionless force $f=FL/k_{\mathrm{B}}T$. The
    symbols correspond to the numerically simulated exact values,
    cf. Eq.~\eqref{eq:current-num}, and the lines to the analytically
    calculated, approximative values,
    cf. Eq.~\eqref{eq:nonlinearmobility}, with the diffusion
    coefficients: $D_{0}=1$ (solid line); $D_{\mathrm{RR}}(x)$ (dashed
    line), cf.
    Eq.~\eqref{eq:diffusionconst}, and $D_{\mathrm{Z}}(x)$
     (dashed-dotted line), cf. Eq.~\eqref{eq:diffusionconst-zwanzig}.
}
\label{fig:comparison}
\end{figure}

Fig.~\ref{fig:comparison} demonstrates that the resumed one dimensional
diffusion coefficient $D_{\text{RR}}(x)$ leads to a considerably better
agreement with the numerical results than the constant diffusion and
the
diffusion coefficient $D_{\text{Z}}(x)$ proposed by Zwanzig.
Therefore we used $D_{\mathrm{RR}}(x)$ for all following calculations.

\section{Validity of the Fick-Jacobs description in the presence
of a constant bias}
\label{sec:validity}

The reduction of dimensionality leading to
the FJ equation relies
on the assumption of equilibration in the transverse direction
which results in an almost uniform distribution of the transversal
positions $y$ at fixed values of the longitudinal coordinate $x$.
One can formulate two different sets of criteria determining first,
whether the FJ equation describes the relaxation towards the
stationary state, say in presence of periodic boundary conditions, or
second, whether the stationary state, but not necessarily the
relaxation towards this state, can be described by the
FJ equation.

Although we here are mainly interested in the second, weaker, type
of criteria which are sufficient to guarantee the validity of the
transport properties predicted by the FJ equation, we shortly formulate
stronger criteria which must be satisfied if the FJ equation is
employed to determine the relaxation towards equilibrium. Then, of
course, the time scale of equilibration in transversal direction
$\tau_{\mathrm{T}}$ must be short compared to the relevant time scales in
longitudinal direction, $\tau_\mathrm{L}$ \cite{Burada_PRE}. The time scale $\tau_\mathrm{T}$ can be
estimated by the time to diffusively cover the widest transversal
distance of the channel. It therefore is given by
\begin{equation}
\tau_\mathrm{T} =2 (a+b)^{2}.
\label{tauT}
\end{equation}
The time scales characterizing the longitudinal motion are the
diffusion time $\tau_{\mathrm{d}L}$ over the length of a period, which is one in our
dimensionless variables, and the time $\tau_{\mathrm{f}L}$
it takes to drag a particle over
this distance by applying the force $f$. These times are given by
\begin{equation}
\tau_{\mathrm{d}L} =1/2 \text{ and } \tau_{\mathrm{f}L} = 1/f
\label{taudf}
\end{equation}
Hence, a necessary condition, that the FJ equation reliably describes
transient processes can be formulated as
\begin{equation}
2 (a+b)^{2} \ll \min (1/2, 1/f)
\label{c1}
\end{equation}
This condition is fulfilled only for rather elongated channels being
at least five times as long as wide.

As
already mentioned in the presence of periodic boundary conditions in
the longitudinal direction, the particle distribution described by the
Fokker-Planck equation \eqref{eq:fp}
approaches a stationary
distribution. Even in the case if the first criterion (\ref{c1}) is
violated the stationary solution of the FJ equation
(\ref{eq:fickjacobs_ours}) may
still yield the correct  marginal
probability density provided transversal
cuts of the two dimensional stationary probability density are
practically constant.
Such a uniform distribution in transversal direction strictly holds
in the absence of externally imposed concentration differences if the
force $f$ vanishes or if
the channel is straight. For channels with varying width the narrow
positions confine the positions of particles. From there they are dragged
by the force and at the same time they perform a diffusive motion
until the channel narrows again. The required uniform distribution in
the transversal direction can only be achieved if the diffusional motion
is fast enough in comparison to the deterministic drift under the
influence of $f$. In other words, the diffusional spreading within the
time the force drags the particle from the narrowest to the widest
place in the channel must be at least of the order of the widest
channel width.  This leads to the second, weaker, criterion
\begin{equation}
(a+b)^{2} \leq \frac{1}{2f}\quad \Leftrightarrow \quad f \leq
\frac{1}{a^{2}} \, \frac{1}{2(1+\kappa)^{2}} \, \propto \frac{1}{a^{2}} .
\label{c2}
\end{equation}
\begin{figure}[t]
  \centering
  \includegraphics{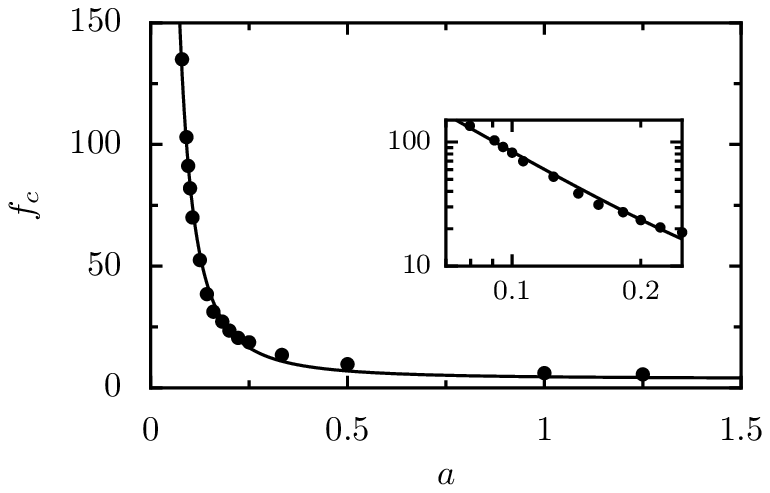}
  \caption{The dependence of the critical value of the dimensionless
    force $f_c$ on the parameter $a$ for  2D channels
    defined by the dimensionless boundary function
    $\omega(x)=a \left(\sin(2\pi x)+1.02 \right)$ is depicted.
    For $f<f_c$ the accuracy of the
    Fick-Jacobs approximation is $\sim 1\%$ (in comparison with the exact
    2D simulational results). The solid line demonstrates the
    $a^{-2}$-dependence of the critical values predicted by Eq.~\eqref{c2}.
    The inset depicts the same data on a  logarithmic scale.}
  \label{fig:critical}
\end{figure}

Eq.~\eqref{c2} provides an estimate of the minimum
forcing above which the FJ  description is expected to fail
in providing an accurate description of the transport properties in
the long time limit.
The quantitative value of the critical
force depends on the level of the prescribed accuracy.
The
criterion demonstrates how the validity of the equilibrium
approximation  depends
on the relevant parameters of the problem and is concordant with that
found for a different scaling in Ref. \cite{Burada_PRE}.

In order to test the accuracy of the FJ  description, we
evaluated the behavior of the nonlinear mobility as a function of
the scaled force $f$, for different values of $a$ according to
Eq.~\eqref{eq:nonlinearmobility} and compared it with
numerical simulations of the corresponding full two dimensional
problem.
The value of the dimensionless force $f$ up to which
the FJ  approximation with spatially dependent diffusion coefficient
$D_{\mathrm{RR}}(x)$ provides an accurate description depends on $a$
just as predicted by Eq.~\eqref{c2}, cf. Fig.~\ref{fig:critical}. 
For large values of $a$ the FJ
equation starts to deviate from the numerically exact behavior already
for rather small forces $f$, whereas for small values of $a$ larger
forces may be applied without violating the FJ equation.

\begin{figure}[t]
\centering
  \includegraphics{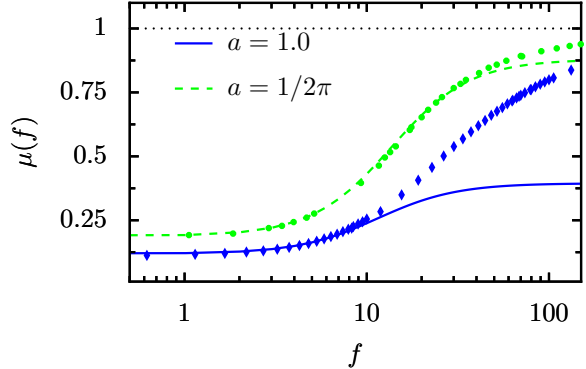}
  \caption{(color online) The numerically simulated (symbols) and
    analytically calculated (cf. Eq.~\eqref{eq:nonlinearmobility}
    -- lines)  dependence of the scaled nonlinear
    mobility $\mu(f)$ {\it vs.} the dimensionless force $f=FL/k_\mathrm{B}T$ is
    depicted for two 2D channel geometries.
    For both channels the scaled half-width is given by
    $\omega(x)= a \left(\sin(2\pi x) + 1.02\right)$;
    $a=1:$ diamonds and solid line(blue),
    $a=1 / (2\pi):$ circles and dashed line (green).
    The dotted line indicates the deterministic limit $\mu(f) =
    {\langle\dot x \rangle}/f = 1$.}
\label{fig:mobility}
\end{figure}

In Fig.~\ref{fig:mobility} the dependence of the nonlinear mobility on
the force $f$ is displayed for two different values of the parameter
$a$ characterizing the geometry of the channel. For the channel with
$a=1$ which, at its widest opening, is approximately four time as wide
as it is long, the
predictions of the equilibration assumptions fails at smaller $f$-values
than in the case of  $a=1/2\pi$.

\section{Transport characteristics: Anomalous temperature dependence and enhancement of diffusion}
\label{trans}

Transport in one dimensional periodic {\it energetic} potentials
behaves very differently from one dimensional periodic systems
with {\it entropic} barriers.
The fundamental difference lies in the temperature dependence of
these models.
Decreasing temperature
in an energetic periodic potential decreases the transition
rates from one period to the neighboring by decreasing the Arrhenius
factor $\exp\{-\Delta V/(k_{B}T)\}$ where $\Delta V$ denotes the
activation energy necessary to proceed by a period \cite{hanggi}.
Hence decreasing temperature leads to a decreasing mobility.
For a one-dimensional periodic system with an entropic potential, a
decrease of temperature leads to an increase of the dimensionless
force parameter $f$ and consequently to an increase of the mobility,
cf. Fig.~\ref{fig:mobility}.

On the other hand, the dependence of the dynamics on the geometry parameter $a$
clearly reflects the entropic effects on the mobility.
A channel with a larger $a$ value has wider openings and therefore
provides more space where the particle can sojourn. This longer
residence time within a period of the channel diminishes the
throughput and consequently the mobility. This is corroborated by the
results of our calculations depicted in
Fig.~\ref{fig:mobility}. For all values of $f$, an increase in value of $a$ leads to a
decrease in the mobility. This holds not only where the FJ equation 
applies but also for large values of $f$ where it fails.

\begin{figure}[t]
\centering
  \includegraphics{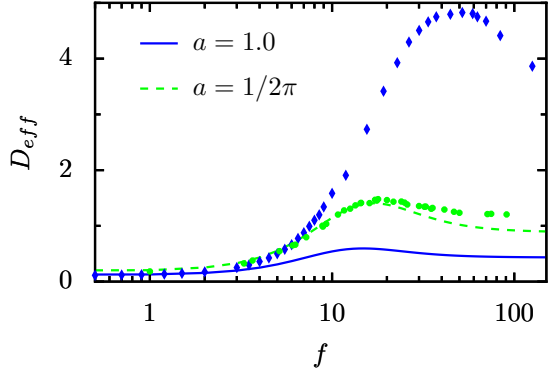}
  \caption{(color online) The numerically simulated (symbols) and
    analytically calculated (cf. Eq.~\eqref{eq:effectivediffusion} --
    lines)  dependence of the effective diffusion coefficient
    $D_{\text{eff}}$ is depicted {\it vs.} the dimensionless force $f=FL/k_\mathrm{B}T$
   for two channels in 2D.
    For both channels the scaled half-width is given by
    $\omega(x)=a\left(\sin(2\pi x) + 1.02 \right)$;
    $a=1:$ diamonds and solid line(blue), $a=1 /(2\pi):$
    circles and dashed line (green).}
\label{fig:diffusion}
\end{figure}

Another interesting effect can be observed for the effective diffusion
if looked as a function of the force $f$. Already the expression for the
effective diffusion \eqref{eq:effectivediffusion} which follows rigorously from the
FJ equation displays a maximum as a function of $f$ which may even
exceed the value 1 of the bare diffusion,
cf. Fig.~\ref{fig:diffusion}. 
For $f\to \infty$ the periodic stationary
distribution approaches a delta function along the $x$ axis and the
effective diffusion approaches the bare value
1. If one decreases the force to finite but still large values then
the stationary distribution acquires a finite width in the transversal
direction with a ``crowded'' region in front of the narrowest place of
the channel Ref.~\cite{Burada_PRE}. The transport becomes more noisy
and consequently the
effective diffusion exceeds the bare value 1. On the other hand if one
starts at $f=0$ the entropic barriers diminish the diffusion such that
the effective diffusion is less than bare diffusion. Consequently,
somewhere in between there must be a value of $f$
with maximal effective diffusion.
For $a=1$ and $b=1.02$
the value of the force at the maximal effective diffusion is outside
the regime of validity of the FJ equation. The numerical simulations
give a much more pronounced peak of the effective diffusion.
For the less entropic
channel with $a=1/2 \pi$ the maximum is at the border of the regime of
validity of the FJ equation, but the enhancement of the effective
diffusion constant is less pronounced than for the larger value $a=1$.
These observations lead us to the conclusion that
entropic effects increase the randomness of transport through a
channel and in this way decrease the mobility and increase the
effective diffusion. {\corrected A similar enhancement of effective
diffusion was found in titled periodic {\itshape energetic} 
potentials \cite{Reimann_PRL, Lindner, Constantini}.}

\section{Conclusions}
\label{conclusions}

In summary, we demonstrated that transport phenomena in periodic
channels with varying width exhibit some features that are radically
different from conventional transport occurring  in energetic
periodic potential landscapes. The most striking difference between
these two physical situations lies in the fact that for a fixed
channel geometry the dynamics is completely characterized by a
single parameter $f= FL/(k_{B}T)$ which combines the external force
$F$ causing a drift, the period length $L$ of the channel, and the thermal
energy $k_{B}T$, which is a measure of the strength of the acting
fluctuating forces. Transport in periodic energetic potentials
depends, at least, on one further parameter which is the height of
the highest barrier separating neighboring periods. This leads to an
opposite temperature dependence of the mobility. While the mobility
of a particle in an energetic potential increases with increasing
temperature the mobility of a particle in a channel of periodically
varying width decreases. The incorporation of the spatial variation
of the channel width as an entropic potential in the FJ equation
allows a qualitative understanding of the dependence of the transport
properties on the channel geometry.

The effective diffusion exhibits a non monotonic dependence versus
 the dimensionless force $f$. It starts out at small $f$ with
a value that is less than the bare diffusion constant, reaches a
maximum with increasing $f$ and finally approaches the value of the
bare diffusion from above.

It is known from the literature that, under certain conditions, the
two-dimensi\-onal Fokker-Planck equation governing the time dependence
of the probability density of a particle in the channel can be
approximated by a one-dimensional Fokker-Planck equation: It is termed
the Fick-Jacobs equation and contains an entropic potential and a
position dependent diffusion coefficient. Various forms of the
diffusion coefficient can be found in the literature. A comparison
of different forms for a particular channel geometry leads to the
conclusion that the expression recently suggested by Reguera and
Rub\'i yields the most favorable agreement. In principle the FJ
equation describes both the transient behavior of a particle and
also the stationary behavior of the particle dynamics which is
approached in the limit of large times, provided appropriate
boundary conditions confining the motion in the direction of the
channel axis are applied. In order to study stationary transport,
periodic boundary conditions must be invoked. We formulated criteria
for the validity of the FJ equation for both the transient and the
stationary regimes and found that the restrictions imposed by the
criterion in the stationary regime are much less serious than those
for the transient dynamics. The estimates, which are based on simple
dynamical arguments, were corroborated by our numerical simulations.

We restricted our analysis to two dimensional channels. A
generalization of the presented methods to three dimensional pores
with varying cross section is in principle straight-forward. We also
confined ourselves to channels with a mirror symmetry about a
vertical axis which in the present case can be chosen  {\corrected at $x=1/4$.
For periodic} channel shapes without this symmetry  ratchet like
transport can be expected even if the unbiased force $f$ of
vanishing temporal average changes periodically in time.

\section*{ACKNOWLEDGEMENTS}
This work has been supported by the DGiCYT under Grant No
BFM2002-01267 (D.R.), ESF STOCHDYN project (G.S., D.R., J.M.R.,
P.H.), the Alexa\-nder von Humboldt Foundation (J.M.R.), the
Volkswagen Foundation (project I/80424, P.H.), the DFG via research
center, SFB-486, project A10 (G.S., P.H.) and via the DFG project
no. 1517/26-1 (P.S.B., P.H.), and by the German Excellence
Initiative via the \textit {Nanosystems Initiative Munich} (NIM) (P.S.B., P.H.).


\end{document}